\documentclass{article}

\usepackage[numbers]{natbib}         
\usepackage[colorlinks]{hyperref}    
\usepackage[english]{babel} 
\usepackage{amssymb}
\usepackage{amsmath}
\usepackage{txfonts}
\usepackage{mathdots}
\usepackage[classicReIm]{kpfonts}
\usepackage[dvips]{graphicx} 
\usepackage[a4paper, portrait, margin=1in]{geometry}
\usepackage{tabularx}
\usepackage{longtable}
\usepackage{multirow}
\usepackage{booktabs}
\usepackage[labelsep=period]{caption}
\usepackage{makecell}
\begin{document}
	\setlength{\parindent}{0pt}
	\setlength{\parskip}{1ex}
	
	\textbf{\Large Learning-Based Synthetic Dual Energy CT Imaging from Single Energy CT for Stopping Power Ratio Calculation in Proton Radiation Therapy}
	
	\bigbreak

	Serdar Charyyev$^+$, Tonghe Wang$^+$, Yang Lei, Beth Ghavidel, Jonathan J. Beitler, Mark McDonald, Walter J. Curran, Tian Liu, Jun Zhou and Xiaofeng Yang*
	
	Department of Radiation Oncology and Winship Cancer Institute, Emory University, Atlanta, GA 30322
	
	$^+$Co-first author

	\bigbreak
	\bigbreak
	\bigbreak

	\textbf{*Corresponding author: }
	
	Xiaofeng Yang, PhD
	
	Department of Radiation Oncology
	
	Emory University School of Medicine
	
	1365 Clifton Road NE
	
	Atlanta, GA 30322
	
	E-mail: xiaofeng.yang@emory.edu

	\bigbreak
	\bigbreak
	\bigbreak
	\bigbreak
	\bigbreak
	\bigbreak

	\textbf{Abstract}

		\textbf{Purpose:} Dual-energy CT (DECT) has been shown to derive stopping power ratio (SPR) map with higher accuracy than conventional single energy CT (SECT) by obtaining the energy dependence of photon interactions. However, DECT is not as widely implemented as SECT in proton radiation therapy simulation. This work presents a learning-based method to synthetize DECT images from SECT for proton radiation therapy.
		
		\textbf{Methods:} The proposed method uses a residual attention generative adversarial network. Residual blocks with attention gates were used to force the model focus on the difference between DECT maps and SECT images. To evaluate the accuracy of the method, we retrospectively investigated 20 head-and-neck cancer patients with both DECT and SECT scans available. The high and low energy CT images acquired from DECT acted as learning targets in the training process for SECT datasets and were evaluated against results from the proposed method using a leave-one-out cross-validation strategy. To evaluate our method in the context of a practical application, we generated SPR maps from sDECT using physics-based dual-energy stoichiometric method and compared the maps to those generated from DECT.
		
		\textbf{Results:} The synthesized DECT images showed an average mean absolute error around 30 Hounsfield Unit (HU) across the whole-body volume. The corresponding SPR maps generated from synthetic DECT showed an average normalized mean square error of about 1\% with reduced noise level and artifacts than those from original DECT.
		
		\textbf{Conclusions:} The accuracy of the synthesized DECT image by our machine-learning-based method was evaluated on head and neck patient, and potential feasibility for proton treatment planning and dose calculation was shown by generating SPR map using the synthesized DECT.
	
	\bigbreak
	\bigbreak
	
	\noindent 
	\section{ INTRODUCTION}
	
	Dual-energy CT (DECT), as an advanced CT imaging scheme, is being introduced into multiple usage in radiology as well as radiation therapy. It has been an important imaging modality with special clinical applications in radiology where conventional single energy CT (SECT) falls short, including bone removal in angiography \cite{RN4, RN1, RN202, RN3}, assessment of myocardial blood supply \cite{RN1185, RN5, RN6}, renal calculi characterization \cite{RN9, RN208}, and diagnosis of gout \cite{RN11, RN10}. In these applications, DECT provides material specific information which is extracted by processing the projection datasets of two different energy spectra acquired by DECT \cite{RN13, RN659, RN15, RN660, RN12}. For example, for head-and-neck cancer patients, the derived iodine concentration maps can help differentiate malignant from nonmalignant lymph nodes and benign posttreatment changes from tumor recurrence, and the virtual non-calcium can be used for detection of bone marrow edema \cite{RN51, RN52}. These material specific images derived from DECT scans are now becoming attractive in radiation therapy (RT) with multiple potential applications proposed: including metal artifact reduction, normal tissue characterization, improved dose calculation, functional imaging for target localization and better organ at risk (OAR) delineation \cite{RN17, RN1597}. 
	
	DECT has been shown to be superior to SECT in proton therapy treatment planning as well. Stopping power ratio (SPR) of protons, the quantity needed to calculate the proton dose distribution and range in today’s treatment planning systems, is currently derived from SECT images by applying a piecewise linear fit between CT Hounsfield Unit (HU) and SPRs, calibrated either on literature data for human tissues or on measurements for tissue substitutes with known properties \cite{RN22, RN47, RN19}. However, the accuracy of SPRs from the empirical fit may be compromised by the patient-specific tissue variations \cite{RN1592} and the ambiguous HU-SPR relationship since some tissues may have the same CT number but different SPRs or vice versa \cite{RN20}. For this reason, methods to estimate the SPR from DECT by utilizing its material differentiation ability has been proposed by several groups \cite{RN25, RN20, RN1592}, among which the physics-based stoichiometric method is widely accepted and used clinically \cite{RN34, RN1592, RN35}. It derives the SPR maps in a voxel-wise manner from DECT image datasets using physical equations, which enables DECT to have the upper hand over SECT in SPR estimation both in phantoms and animal tissue studies \cite{RN27, RN28, RN3010}. 
	
	Despite all the mentioned potentials, DECT did not enjoy the widespread implementation in the clinics as much as SECT did mainly due to the costs associated with acquiring and operating DECT and declining reimbursement \cite{RN29}. Moreover, the sub-optimal implementations of DECT acquisition would lead to extra noise and artifacts in addition to those present in SECT. For example, the two-sequential-scan scheme is greatly limited by its poor temporal coherence due to the time interval between two scans \cite{RN38, RN37}. Other implementations such as low/fast kVp-switching also has limitations in lower spectral resolution, slow gantry motion and less projections per spectrum \cite{RN38}. Dual-source CT has orthogonal setup of two x-ray source and detectors pairs, which is inherently prone to cross scatter radiation between noncorresponding orthogonal detector rows \cite{RN53, RN38, RN54}, although it can be mitigated by additional software and hardware approaches \cite{RN56, RN1593}. Dual-source CT has cross scatter between the two source-detector pairs. These non-idealities would downgrade the DECT image quality, and would be significantly magnified in the derived results such as SPR due to the ill-conditioned property of material differentiation process \cite{RN39, RN1685, RN41} in image domain. TwinBeam CT has degraded accuracy in material differentiation cause by decreased energy spectra separation when compared with other DECT modalities \cite{RN1343, RN43} . The strong overlapping of energy spectra of linear attenuation coefficient among different materials would aggravate the noise magnification from DECT dataset to its derived results. 
	
	Considering limited availability of DECT and the above limitations in the current DECT implementation, using SECT to generate synthetic DECT (sDECT) and calculating SPR from sDECT is an alternative approach of particular clinical and research interest for proton therapy treatment planning. Inspired by the recent advances in machine learning, a few studies proposed to synthesize DECT datasets using deep learning framework \cite{RN45, RN44}. These studies mainly focus on diagnostic purposes, while the accuracy of the derived SPR has not been studied.
	
	In this work, we developed a deep learning-based method to synthesize DECT images, i.e. synthetic low energy CT (sLECT) image and synthetic high energy CT (sHECT) image, from SECT images. Compared with other machine-learning based methods, the advantages include adding three discriminators to enhance the reality of synthetic DECT and adding gradient difference error in loss function to increase the similarity of the local tissue structure between sDECT and its training target. We retrospectively investigated 20 head-and-neck patients with TwinBeam DECT scan, and its composed SECT scan acquired during proton therapy treatment planning. These acquired DECT images, i.e. high energy CT (HECT) and low energy CT (LECT), acted as learning targets in the training process for SECT images, and were evaluated against results from the proposed method using a leave-one-out cross-validation strategy. The accuracy of sDECT obtained by the proposed method was quantified with multiple image quality metrics. To evaluate our method in the context of a practical application, we generated SPR maps from sDECT and compared it to those from DECT. We are not aiming to improve SPR accuracy over real DECT, but to have a comparable SPR with real DECT. SPR generation is one of many applications where sDECT can be used in clinic. The proposed learning-based method for sDECT can thus be potentially useful in more clinical workflows.

	\noindent 
	\section{Methods}
	\noindent 
	\subsection{Image acquisition}
	
	In this retrospective study, we analyzed the dataset of 20 patients with squamous cell carcinoma in head-and-neck region. Patient selection standard is head-and-neck patients who were scanned in TwinBeam CT with DECT mode. The 20 patients included 13 males and 7 females with ages ranging from 25 to 89. Their tumor sites vary from patient to patient including larynx, buccal mucosa, tongue, etc., and 12 patients underwent excisions. Each patient had CT simulation by TwinBeam CT in DECT mode with 110 s delay after 100 mL Omnipaque 300 iodine contrast injected at 2.5 mL/s, followed by treatment planning for proton radiation therapy. Institutional review board approval was obtained with no informed consent required for this HIPAA-compliant retrospective analysis. 
	
	The DECT images were acquired using a Siemens SOMATOM Definition Edge TwinBeam CT scanner at 120 kVp with the patient in treatment position (pitch: 0.45, rotation time: 0.5 s, scan time: around 30 s, CTDIvol: around 20 mGy, reconstruction kernel: Q30f, tube current ranges from 500 to 650 mA, and metal artifact correction was in use). The 120 kVp X-rays were split into high and low spectra by 0.05 mm gold and 0.6 mm tin filters, yielding high energy and low energy scans, respectively. Meanwhile, composed polychromatic single energy images were reconstructed from raw projection dataset by disregarding spectral differences, i.e. it is reconstructed as conventional single energy CT before any dual energy-related process \cite{RN1597}. Note that it is not a derivation from DECT. The composed images served as the treatment planning CT for tumor/organs-at-risk (OAR) delineation and dose calculation in clinic, and acted as the SECT in this study. All HECT/LECT/SECT images were reconstructed by Siemens Syngo CT VA48A with spacing 1.27 × 1.27 × 1.5 mm$^3$.
	 
	\noindent 
	\subsection{Synthetic low and high energy CT generation}
	\subsubsection{Workflow}
	
	Figure 1 outlines the schematic workflow of our prediction method, which includes a training and a synthesis stage. During training, for a given SECT and its corresponding LECT and HECT images, the LECT and the HECT images were used as the learning-based regression target of the SECT image. 3D patches with size of 96×96×64 were extracted from SECT image via sliding a window from those images with an overlap between two neighboring patches. The overlap size was set to 72×72×40 in order to obtain a large training data set. To further enhance the variance of the training data, data augmentation, which includes rotation, flipping, scaling and rigid warp, was applied to the patches of training data. Then, these patches were fed into a 3D deep learning framework, a generative adversarial network (GAN)-based framework. The patches were first fed into a generator as single-modality input to obtain an equal sized two channel outputs. The first channel output was regarded as sHECT patches. The second channel output was regarded as sLECT patches.
	
	In order to learn the detailed differences between SECT and HECT/LECT images, we used an attention residual U-Net to implement the network architecture of the generator. Then, three discriminators were used to judge the realism of the sHECT, sLECT, and subtraction between sHECT and sLECT images. Each discriminator, implemented by a fully convolution network (FCN)-based network architecture, was used to discriminate the real one (obtained from ground truth HECT, or LECT, or subtraction between ground truth HECT and LECT) from the synthetic one, and thus to enhance the realism of the sHECT and sLECT images. Based on the discriminator’s results, the difference image’s realism between HECT and LECT images would be improved. After training, the patches were extracted from a new arrival patient’s SECT image and were fed into the trained generator to obtain the sHECT and sLECT patches. Finally, by using averaging-based patch fusion, the sHECT and sLECT images of the new arrival patient were obtained.

	\begin{figure}
		\begin{center}
		\noindent \includegraphics*[width=6.50in, height=4.20in, keepaspectratio=true]{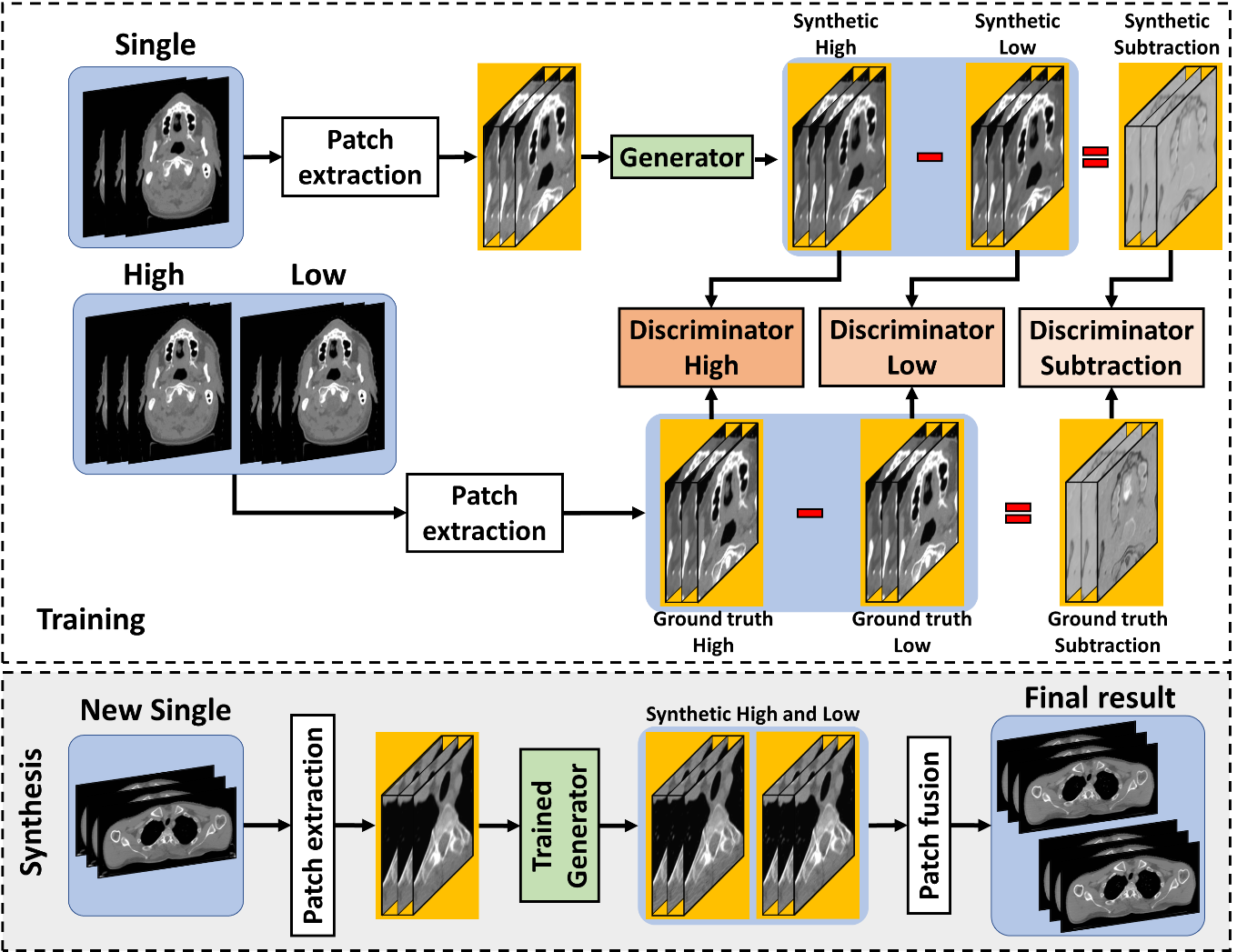}
		\end{center}
		
		\noindent Fig. 1. The systematic flowchart of the proposed method. The first row shows the training stage. The second row shows the prediction stage.
	\end{figure}

	\noindent 
	\subsubsection{Network architecture}
	
	Figure 2 shows the network architecture of the generator and discriminators used in our proposed method. As shown in the figure the generator architecture is an end-to-end attention residual U-Net and is composed of an encoding and a decoding path. The encoding path is composed of three convolution layers with stride size of two to reduce the feature maps’ size and several convolution layers with stride size of one. The decoding path is composed of three deconvolution layers to obtain end-to-end mapping, several convolution layers and a tanh layer to perform the regression. Nine short residual blocks were used as short skip connections between encoding and decoding paths. A long residual block was used as a long skip connection, which bypasses the feature maps from the first convolution layer to the last convolution layer, to enforce all the hidden layers of the generator focusing on learning the difference between input SECT image and the output HECT and LECT images. Three attention gates were integrated into the long skip connections to forward and highlight the informative features from encoding path. 
	
	Through the short residual block, an identity (input) bypasses the hidden layers of a residual block, thus these hidden layers are enforced to learn differences between SECT and HECT/LECT images in a deep level. A short residual block is implemented by two convolution layers within residual connection and an element-wise sum operator \cite{RN30}.
	
	Attention gates were integrated into long skip connection of generator architecture to capture the most relevant semantic contextual information without enlarging the receptive field \cite{RN31}. The feature maps extracted from the coarse scale were used in gating to disambiguate irrelevant and noisy responses in long skip connections. This was performed immediately prior to the concatenation operation to merge only relevant activations. Additionally, attention gates filter the neuron activations during both the forward pass and the backward pass. 
	
	As can be seen from Figure 2, the network architecture of discriminator is a traditional FCN \cite{RN32}. In clinic, a lot of useful information as mentioned in the introduction section is derived from the difference image between HECT and LECT images. Therefore, in addition to using two separate discriminators to judge the realism of sHECT and sLECT images, a third discriminator judging the subtraction image between sHECT and sLECT was also utilized. This discriminator takes the subtraction between ground truth HECT and LECT images as the real one.
	
	\begin{figure}
		\begin{center}
			\noindent \includegraphics*[width=6.50in, height=4.20in, keepaspectratio=true]{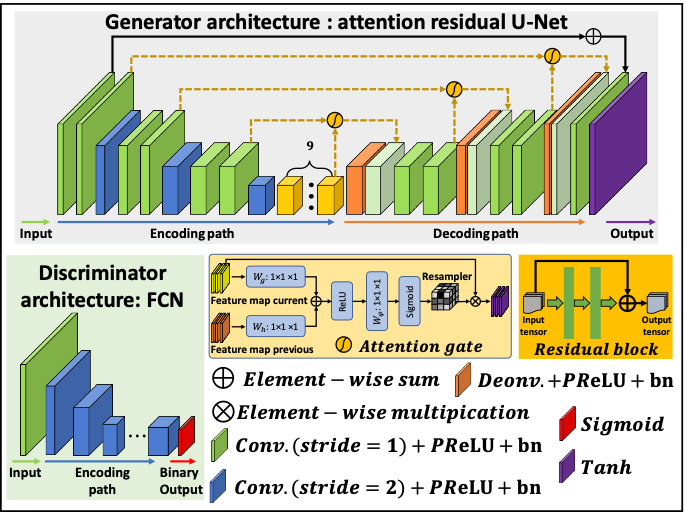}
		\end{center}
		
		\noindent Fig. 2. The network architectures of both generator and discriminator.
	\end{figure}
	
	\noindent 
	\subsubsection{Loss function}
	
	As described above, the network relies on continuous improvement of a generator network and three discriminator networks. The accuracy of both networks is directly dependent on the design of their corresponding loss functions. In this work, mean absolute error (MAE) and gradient difference error (GDE) \cite{RN32} were used as a compound loss to calculate the loss function of generator. MAE is the magnitude of the difference between the ground truth image and the evaluated image and is given by:
	\begin{equation} 
	MAE = \left|\left|I-I_{0}\right|\right|_{1}/N,                      
	\end{equation}
	where $I_0$ is the vector of pixels in ground truth HECT, LECT or the difference image of the those two, and $I$ is the vector of pixels in sHECT, sLECT or difference image of the two, and $N$ is the total number of pixels in calculation. $\left|\left|\bullet\right|\right|_{1}$ indicates L1-norm. The MAE loss forces the generator to synthesize HECT and LECT images with accurate voxel intensity to a level of ground truth HECT and LECT images. Between any two images, the GDE is defined as:
	\begin{equation} 
	\begin{split}
	GDE(Z,Y)=\sum_{i,j,k}\left\{\begin{matrix}\left(\left|Z_{i,j,k}-Z_{i-1,j,k}\right|-\left|Y_{i,j,k}-Y_{i-1,j,k}\right|\right)^2 \\+\left(\left|Z_{i,j,k}-Z_{i,j-1,k}\right|-\left|Y_{i,j,k}-Y_{i,j-1,k}\right|\right)^2\ \\+\ \left(\left|Z_{i,j,k}-Z_{i,j,k-1}\right|-\left|Y_{i,j,k}-Y_{i,j,k-1}\right|\right)^2\\\end{matrix}\right\}  
	\end{split}                       
	\end{equation}
	where Z and Y are any two images, and i, j, and k represent pixels in x, y, and z. The GDE loss forces the sHECT and sLECT images’ gradient structure to a level of ground truth HECT and LECT images. This compound loss function is defined as follows:
	\begin{equation} 
	\begin{split}
	L_g\left(G,I_{Single}\right)=\begin{matrix}\begin{matrix}MAE\left(G_{High}\left(I_{Single}\right),I_{High}\right)+MAE\left(G_{Low}\left(I_{Single}\right),I_{Low}\right)\\+MAE\left(G_{High}\left(I_{Single}\right)-G_{Low}\left(I_{Single}\right),I_{High}-I_{Low}\right)\\\end{matrix}\\+\lambda\bullet\left\{\begin{matrix}GDE\left(G_{High}\left(I_{Single}\right),I_{High}\right)+GDE\left(G_{Low}\left(I_{Single}\right),I_{Low}\right)\\+GDE\left(G_{High}\left(I_{Single}\right)-G_{Low}\left(I_{Single}\right),I_{High}-I_{Low}\right)\\\end{matrix}\right\}\\\end{matrix} 
	\end{split}                       
	\end{equation}
	where $G_{High}\left(I_{Single}\right)$ denotes the sHECT image via feeding SECT image $I_{Single}$ into the generator G and fusing the patches from first channel outputs, $G_{Low}\left(I_{Single}\right)$ denotes the sLECT image via feeding SECT image  $I_{Single}$ into the generator G and fusing the patches from second channel outputs. $\lambda$ is determined via 5-fold cross-validation on training dataset and was set as 0.5. The compound loss function takes the error of HECT, LECT, and the subtraction between HECT and LECT into account.
	
	In addition to compound generator loss, the other goal of generator is to produce the synthetic images that can fool the discriminators via minimizing adversarial losses, which relies on the output of the discriminators, i.e., the distribution of feeding sHECT image into the discriminator of HECT image, the distribution of feeding sLECT image into the discriminator of LECT image, and the distribution of feeding subtraction image between sHECT and sLECT images into the discriminator of subtraction images. Thus, the adversarial loss of generator is defined by:
	\begin{equation} 
	\begin{split}
	L_{adv}\left(G,D_{High},D_{Low},D_{Sub},I_{Single}\right)=\begin{matrix}SCE\left[D_{High}\left(G_{High}\left(I_{Single}\right)\right),1\right]\\+SCE\left[D_{Low}\left(G_{Low}\left(I_{Single}\right)\right),1\right]\\+SCE\left[D_{Sub}\left(G_{High}\left(I_{Single}\right)-G_{Low}\left(I_{Single}\right)\right),1\right]\\\end{matrix}
	\end{split}                       
	\end{equation}
	where $D_{High}$ denotes the discriminator of HECT, $D_{Low}$ denotes the discriminator of LECT, $D_{Sub}$ denotes the discriminator of subtraction images. The function $SCE\left(\bullet,1\right)$ is the sigmoid cross entropy between the input image and a unit mask. 
	
	Finally, the optimization of generator is obtained by minimizing the compound loss and adversarial loss as follows:
	\begin{equation} 
	\begin{split}
	G={arg\min}_G\left\{L_g\left(G,I_{Single}\right)+\mu L_{adv}\left(G,D_{High},D_{Low},D_{Sub},I_{Single}\right)\right\}.
	\end{split}                       
	\end{equation}
	$\mu$ is determined via 5-fold cross-validation on training dataset and was set as 0.1. In order to improve the discriminating ability, which can well classify the synthetic one from real one, the optimization of decimators is obtained as follows:
	\begin{equation} 
	\begin{split}
	D_{High}={\rm argmin}_{D_{High}}{\left\{SCE\left[D_{High}\left(G_{High}\left(I_{Single}\right)\right),0\right]+\ SCE\left[D_{High}\left(I_{High}\right),1\right]\right\}},
	\end{split}                       
	\end{equation}
	
	\begin{equation} 
	\begin{split}
	D_{Low}={\rm argmin}_{D_{Low}}{\left\{SCE\left[D_{Low}\left(G_{Low}\left(I_{Single}\right)\right),0\right]+\ SCE\left[D_{Low}\left(I_{Low}\right),1\right]\right\}},
	\end{split}                       
	\end{equation}
	
	\begin{equation} 
	\begin{split}
	D_{Sub}={\rm argmin}_{D_{Sub}}{\left\{SCE\left[D_{Sub}\left(G_{High}\left(I_{Single}\right)-G_{Low}\left(I_{Single}\right)\right),0\right]+\ SCE\left[D_{Sub}\left({I_{High}-I}_{Low}\right),1\right]\right\}}.
	\end{split}                       
	\end{equation}
	
	\noindent 
	\subsection{Evaluation}
	
	The cohort of 20 patients was used to evaluate our method using leave-one-out cross-validation. For one test patient, the model is trained by the remaining 19 patients. This train-and-test procedure is repeated 20 times to let each patient be used as a test patient exactly once. For each training and test procedure, the model is initialized and re-trained for next test patient by training another group of 19 patients. The training datasets and testing datasets are separated and independent during each study. For our training, we used data augmentation and 3D patch-based method to increase training data variation. Flipping, rotation, scaling and rigid warping were used to enlarge the data size by 72 times. Patch size was set to 96x96x32.
	
	To quantify the prediction quality of synthetic DECT images, three commonly used metrics were applied, including MAE, peak signal-to-noise ratio (PSNR), and normalized cross correlation (NCC). All comparison metrics are calculated within patient body for each patient.
	
	PSNR is the ratio between the maximum possible power of a signal and the power of corrupting noise that affects the fidelity of its representation and is calculated as follows:
	
	\begin{equation} 
	PSNR=10\ \times\log_{10}{\left(\frac{MAX^2}{\left|\left|I-I_{0}\right|\right|_{2}^{2}}\right)}                  
	\end{equation}
	where MAX is the maximum pixel intensity in $I_0$ and $I$, and $\left|\left|\bullet\right|\right|_{2}$ indicates L2-norm. 
	
	The NCC is a measure of the similarity of image structures and is used commonly in pattern matching and image analysis \cite{RN33}:
	\begin{equation}
		NCC=\frac{cov(I_{0},I)}{\sigma_{I_{0}}\sigma_{I}},                
	\end{equation}
	where $cov(\bullet,\bullet)$ indicates covariance, $\sigma_{I_{0}}$ and $\sigma_{I}$ are the standard deviation of ground truth and predicted results, respectively.
	
	To compare the SPR maps generated by the ground truth DECT and sDECT, physics-based dual-energy stoichiometric method was implemented. This method assumes that the effective atomic number of a material can be expressed in an equation as the ratio of its image intensities between low and high energy CT images with empirical parameters. To determine these empirical parameters, phantom with multiple material rods of known material composition and therefore effective atomic number are scanned and measured for the corresponding CT number ratio at different energy levels. With these parameters determined, the equation can then be used to calculate the effective atomic number for unknown material with its measured CT number ratio of dual energies. The SPR is calculated from the effective atomic number using Bethe-Bloch equation. 
	
	We used the Gammex RMI 467 electron density phantom with the chemical compositions of inserted materials specified by the manufacturer. This phantom has limitations as its size is bigger than human head, however, in order to include an appropriate number of material rods, a phantom with such size is necessary. Moreover, the phantom is short; therefore, we skipped several slices at the first and last axial slices when we measured on the images. The SPR value of each rod was calculated from the known chemical compositions using equations reported \cite{RN47}. The rods were randomly placed throughout the phantom for 5 different scans so an average HU value, relatively independent of positioning, could be obtained. The same CT scanning protocol as used for patient scans was used during the phantom scan. The HU was measured on each rod at both high and low energy images, and calibrated with the calculated SPR values using equations reported \cite{RN48}. The calibration was then applied on patient DECT images to generate SPR maps.
	
	The overall accuracy of SPR maps generated by the sDECT is quantified by the PSNR and NCC as defined above with those of DECT as ground truth, as well as by normalized mean square error (NMSE) which measures the average error among all pixels. NMSE is defined as \cite{RN1821},
	\begin{equation}
	NMSE=\frac{\left|\left|SPR-SPR_{0}\right|\right|_{2}^{2}}{\left|\left|SPR\right|\right|_{2}^{2}},                
	\end{equation}
	where $SPR_{0}$ and $SPR$ are the SPR maps generated by DECT and sDECT, respectively. To quantify the SPR quality in different materials, we classified the patient anatomy into four regions of interest based on HU numbers on SECT: lung [-900 -300], adipose [-300 0], soft tissue [0 300], and bone [300, 2000]. Mean error is used to quantify the bias of SPR in each volume of interest (VOI), which is the difference of mean SPR value,
	\begin{equation}
	ME\left(j\right)=\frac{\sum_{i\in ROI_j}{SPR\left(i\right)-S{PR}_0\left(i\right)}}{\sum_{i\in ROI_j}{S{PR}_0\left(i\right)}},                
	\end{equation}
	where i indicates the index of the ith pixel in the jth VOI. Noise within each VOI is quantified by the standard deviation of SPR values of that VOI, and artifacts are quantified by the percentage of voxels that deviate by 3 standard deviations from mean.
	
	Our proposed method was compared to other investigated DECT prediction methods proposed by by Zhao et al. (Method A) \cite{RN44} and Li et al. (Method B) \cite{RN45}, both of which are based on convolutional neural networks (CNN). 
	
	\noindent 
	\section{Results}
	
	Figure 3 summarizes the results of our proposed DECT prediction method on an exemplary case. Ground truth (column 1) and synthetic images (column 2) are not visually discernible without detailed analysis. The subtraction image of HECT and sHECT or LECT and sLECT (column 3) shows this in more detail. From column 3 it is visible that large difference of the ground truth and synthetic image is mostly at the boundary of patient. Row 3 shows the subtraction image of HECT and LECT or sHECT and sLECT which are very similar, except for subtraction image of sHECT and sLECT is smoother because we introduced a lot of convolution layers. Columns 4-6, the magnified images of the region of interest indicated by yellow rectangle selection, show that the prediction algorithm can synthesize finely detailed image structures (column 5) of the ground truth HECT and LECT (column 4) images.

	\begin{figure}
		\begin{center}
		\noindent \includegraphics*[width=6.50in, height=4.20in, keepaspectratio=true]{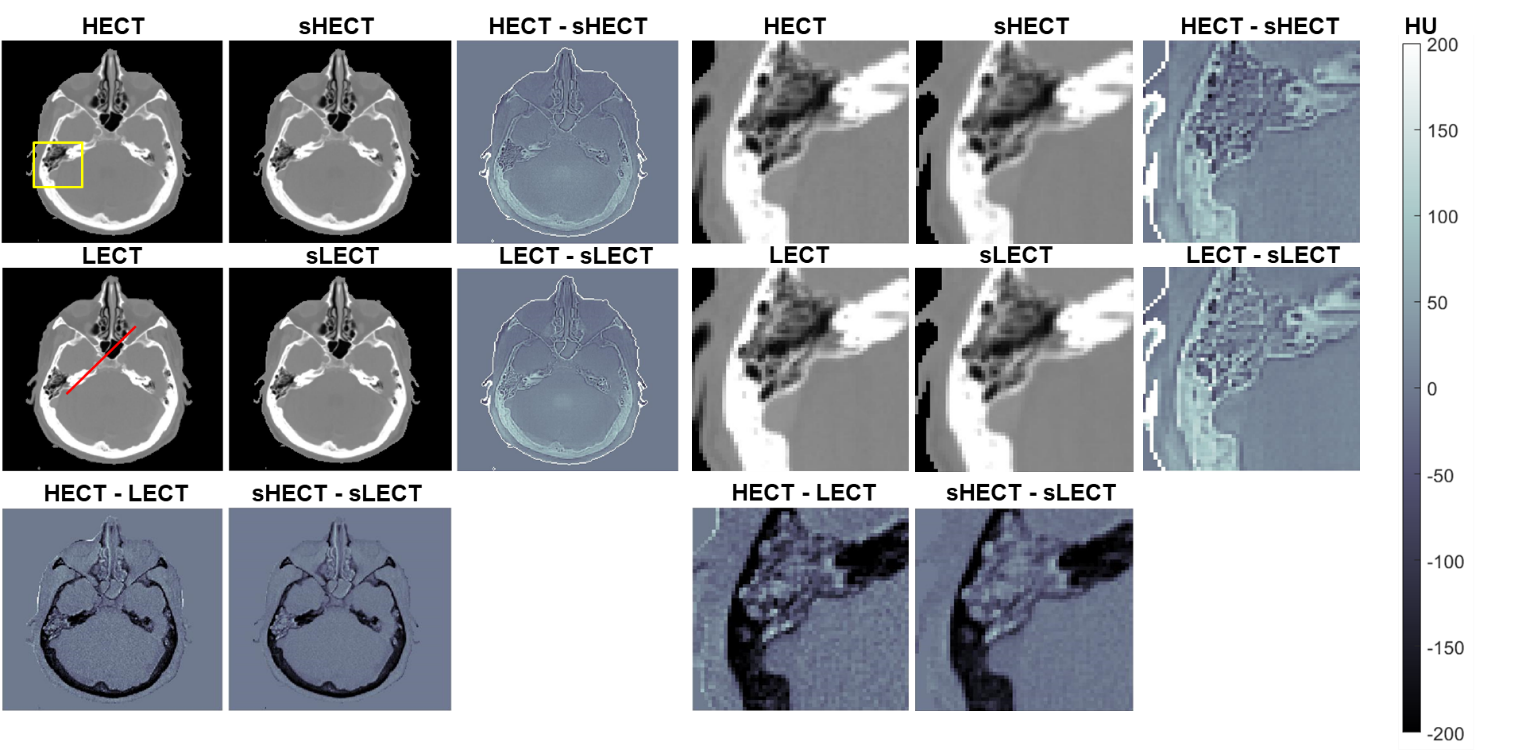}
		\end{center}
		
		\noindent Fig.3 Results of our proposed DECT prediction method in one patient. The first column shows the ground truth HECT, LECT and the difference between those two in rows one, two and three respectively. The second column shows the synthetic HECT, LECT and the difference between those two in rows one, two and three respectively. The third column shows the difference between HECT and sHECT in row one and the difference between LECT and sLECT in row two. Columns 4-6 are the magnified images of the region of interest indicated by yellow rectangle selection. CT display window: [-1000 1000].
	\end{figure}

	\begin{figure}
		\begin{center}
			\noindent \includegraphics*[width=6.50in, height=4.20in, keepaspectratio=true]{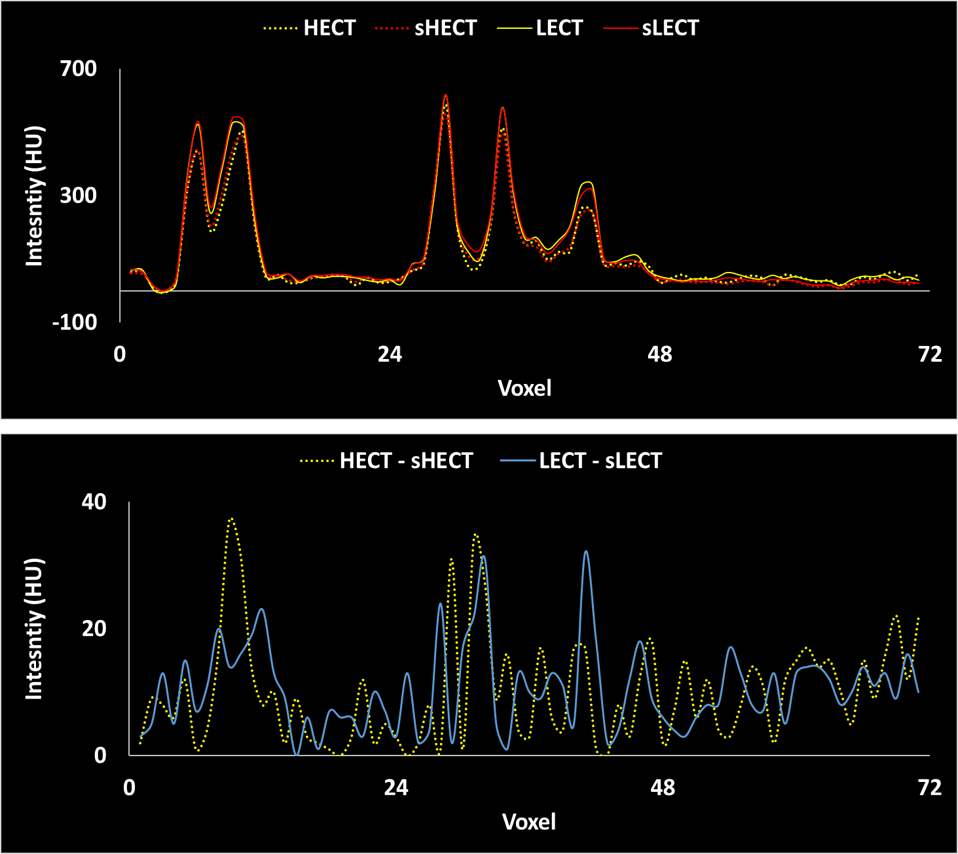}
		\end{center}
		
		\noindent Fig.4 The line profile, (A: for HECT, sHECT, LECT and sLECT), (B: for HECT – sHECT and LECT – sLECT), through the axial images shown in Figure 3 corresponding to the red line drawn on the LECT image (at column one and row two).
	\end{figure}

	In profile (red line in Figure 3) analysis, our deep learning-based method has shown promising results with small HU difference (mostly <20 HU as can be seen from Figure 4B) and similar HU profile across regions with rapid HU change. As illustrated in Figure 4A, profiles of HECT image, dotted yellow line, and sHECT image, dotted red line, are overlapping as do profiles of LECT image, solid yellow line, and sLECT image, solid red line. The former shows difference from the latter at higher HU numbers as expected. 
	
	Similar results as in Figure 3 were obtained in the lung region for other two test patients as shown in the sagittal (left 2 columns) and coronal (right 2 columns) views in the Figure 5. Performance of the proposed method in contrast-enhanced thyroid is shown in Figure 6A alongside with line profile (red line in Figure 6A) through thyroid, Figure 6B.
	
	The quantitative results for all test cases are summarized in Table 1. The mean ± standard deviation (SD) of MAE, PSNR and NCC for sHECT image are 30.23 ± 4.02 HU, 31.83 ± 0.98 dB, and 0.97 ± 0.01, respectively. The mean±SD of MAE, PSNR and NCC for sLECT image are 28.57 ± 5.03 HU, 32.02 ± 1.11 dB, and 0.97 ± 0.01, respectively. The mean±SD of MAE, PSNR and NCC for the subtraction image of sHECT and sLECT are 16.05 ± 1.91 HU, 33.19 ± 3.18 dB, and 0.95 ± 0.008, respectively.
	
	\begin{figure}
		\begin{center}
			\noindent \includegraphics*[width=6.50in, height=4.20in, keepaspectratio=true]{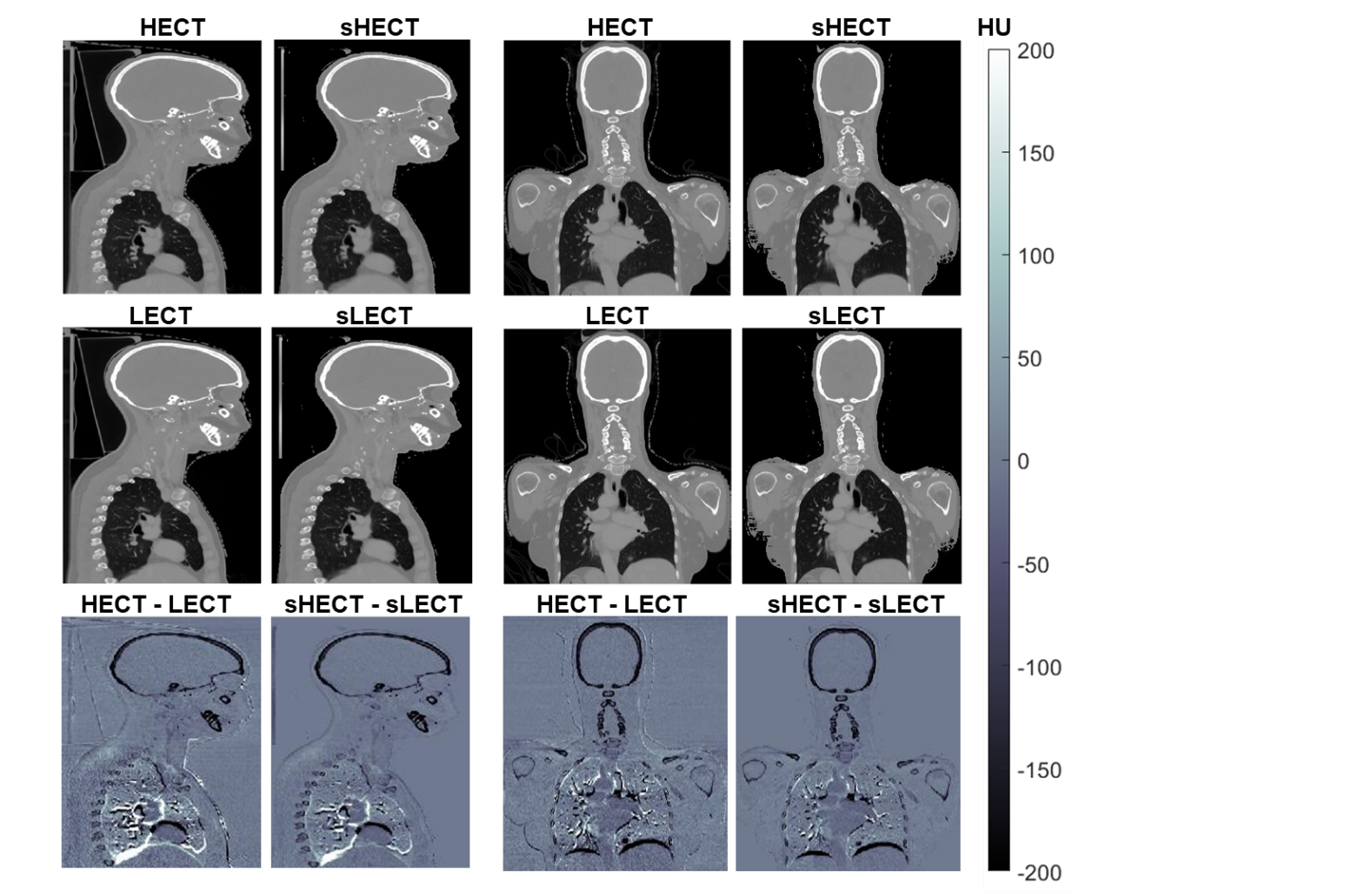}
		\end{center}
		
		\noindent Fig.5 Summary figure of results obtained for two other patients in the lung region in the sagittal and coronal views. 
	\end{figure}

	\begin{figure}
		\begin{center}
			\noindent \includegraphics*[width=6.50in, height=4.20in, keepaspectratio=true]{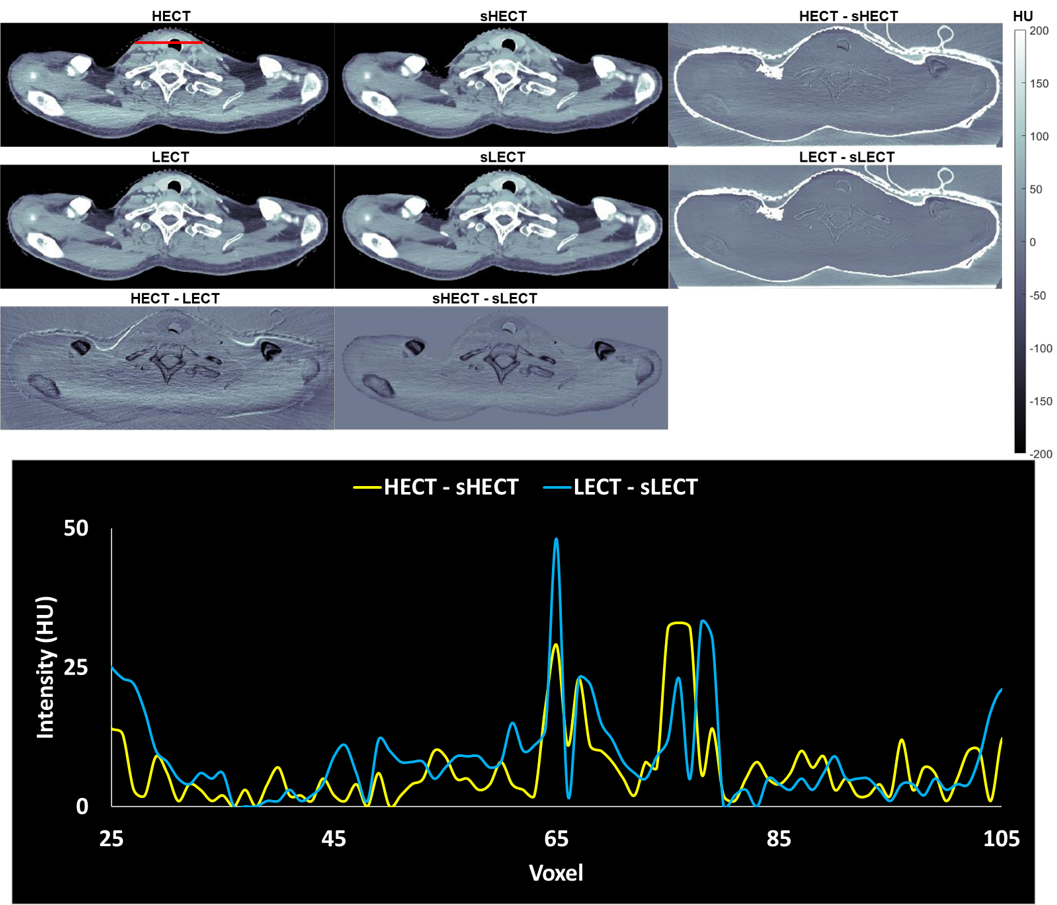}
		\end{center}
		
		\noindent Fig.6 (A) Summary figure of results obtained for contrast-enhanced tissue, thyroid. (B) Line profile through contrast-enhanced tissue. 
	\end{figure}
	
	\begin{table}[htbp]
		\centering
		\caption{Statistics of synthetic DECT for the MAE, PSNR and NCC values for all patients.}
		\hspace*{-1cm}
		\begin{tabular}{ccccc}
			\hline
			\multicolumn{2}{c}{}                & sHECT      & sLECT      & sHECT-sLECT \\ \hline
			\multirow{4}{*}{MAE(HU)}  & Mean±SD & 30.23±4.02 & 28.57±5.03 & 16.05±1.91  \\ \cline{2-5} 
			& Median  & 29.91      & 28.06      & 16.16       \\ \cline{2-5} 
			& Min     & 24.34      & 20.55      & 11.94       \\ \cline{2-5} 
			& Max     & 40.01      & 37.82      & 20.66       \\ \hline
			\multirow{4}{*}{PSNR(dB)} & Mean±SD & 31.83±0.98 & 32.02±1.11 & 33.19±3.18  \\ \cline{2-5} 
			& Median  & 31.79      & 32.01      & 32.79       \\ \cline{2-5} 
			& Min     & 29.64      & 29.61      & 28.80       \\ \cline{2-5} 
			& Max     & 33.87      & 34.05      & 40.41       \\ \hline
			\multirow{4}{*}{NCC}      & Mean±SD & 0.97±0.01  & 0.97±0.01  & 0.95±0.008  \\ \cline{2-5} 
			& Median  & 0.97       & 0.98       & 0.95        \\ \cline{2-5} 
			& Min     & 0.94       & 0.95       & 0.93        \\ \cline{2-5} 
			& Max     & 0.98       & 0.99       & 0.96        \\ \hline
		\end{tabular}
	\end{table}%

	 Figure 7 shows the results of our proposed DECT prediction method as compared to Method A and Method B. Table 2 is a comparison of proposed method to method A and method B with MAE and ME for four different regions of interest: bone, fat, soft tissue and lung. Our proposed method was able to generate similar, if not superior, quality as compared to existing methods.
	 
	 \begin{figure}
	 	\begin{center}
	 		\noindent \includegraphics*[width=6.50in, height=4.20in, keepaspectratio=true]{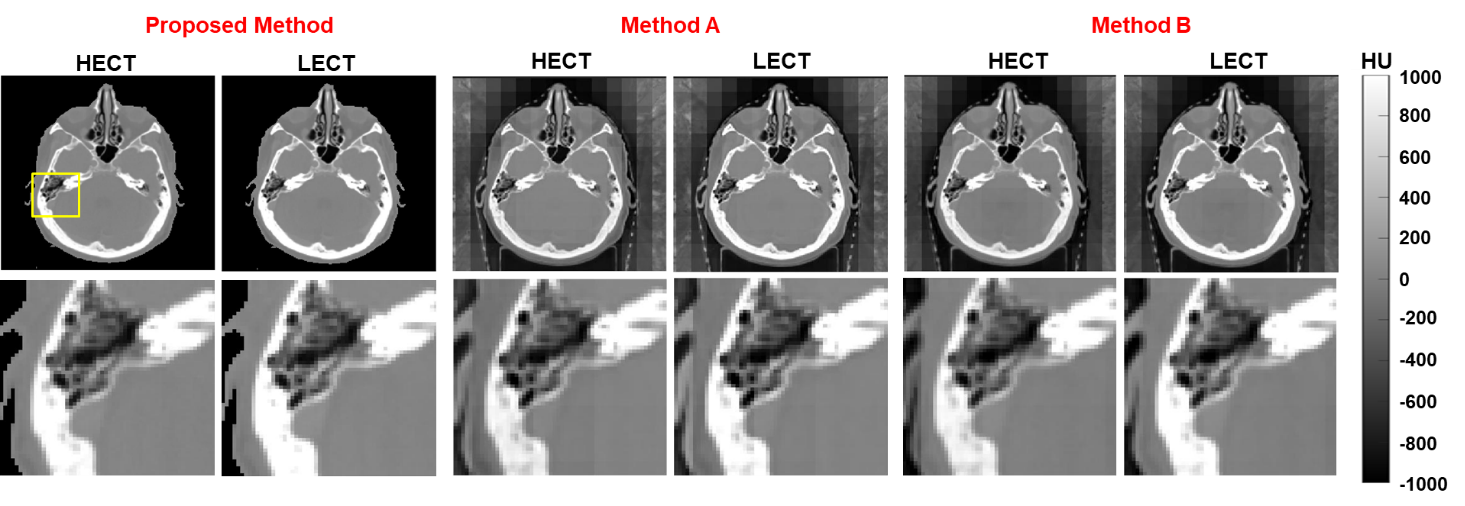}
	 	\end{center}
	 	
	 	\noindent Fig.7 Results of our proposed DECT prediction method compared to Method A and Method B. 
	 \end{figure}
 
 	\begin{table}[htbp]
 		\centering
 		\caption{Comparison of proposed method to method A and method B with MAE and ME for four different regions of interest: bone, fat, soft tissue and lung.}
 		\hspace*{-1cm}
 		\begin{tabular}{cccccccc}
 			\hline
 			\multicolumn{2}{c}{\multirow{2}{*}{}}                                                 & \multicolumn{2}{c}{Proposed Method} & \multicolumn{2}{c}{Method A}  & \multicolumn{2}{c}{Method B}  \\ \cline{3-8} 
 			\multicolumn{2}{c}{}                                                                  & MAE (HU)         & ME (HU)          & MAE (HU)      & ME (HU)       & MAE (HU)      & ME (HU)       \\ \hline
 			\multirow{4}{*}{sHECT}                                                  & Bone        & 34.85 ± 10.04    & 19.00 ± 10.63    & 90.41 ± 20.05 & 67.92 ± 21.67 & 70.49 ± 14.31 & 45.28 ± 16.55 \\ \cline{2-8} 
 			& Fat         & 33.24 ± 6.71     & 8.58 ± 7.19      & 77.00 ± 16.83 & 26.87 ± 22.92 & 69.72 ± 15.97 & 28.55 ± 23.61 \\ \cline{2-8} 
 			& Soft Tissue & 26.98 ± 3.71     & 13.95 ± 6.29     & 59.28 ± 11.00 & 34.38 ± 16.37 & 53.20 ± 11.35 & 34.07 ± 17.87 \\ \cline{2-8} 
 			& Lung        & 18.58 ± 3.78     & 6.97 ± 3.86      & 25.44 ± 4.86  & 5.62 ± 3.11   & 27.61 ± 4.50  & 9.96 ± 4.79   \\ \hline
 			\multirow{4}{*}{sLECT}                                                  & Bone        & 37.49 ± 10.85    & 22.57 ± 11.22    & 91.18 ± 20.86 & 62.36 ± 21.28 & 72.71 ± 15.07 & 44.61 ± 17.94 \\ \cline{2-8} 
 			& Fat         & 33.46 ± 7.66     & 8.50 ± 6.49      & 74.07 ± 17.47 & 30.41 ± 24.79 & 70.43 ± 16.56 & 23.92 ± 21.23 \\ \cline{2-8} 
 			& Soft Tissue & 26.87 ± 4.46     & 11.84 ± 7.16     & 60.34 ± 12.14 & 39.94 ± 17.77 & 52.91 ± 11.57 & 31.75 ± 17.34 \\ \cline{2-8} 
 			& Lung        & 17.72 ± 4.07     & 7.68 ± 4.16      & 23.90 ± 4.89  & 5.84 ± 3.30   & 26.86 ± 4.86  & 11.24 ± 5.08  \\ \hline
 			\multirow{4}{*}{\begin{tabular}[c]{@{}c@{}}sHECT\\ -sLECT\end{tabular}} & Bone        & 21.01 ± 3.57     & 3.58 ± 2.51      & 34.16 ± 3.36  & 7.48 ± 5.11   & 29.03 ± 4.08  & 2.25 ± 1.56   \\ \cline{2-8} 
 			& Fat         & 12.27 ± 1.59     & 2.56 ± 2.16      & 20.43 ± 3.26  & 4.35 ± 3.15   & 21.13 ± 3.27  & 6.60 ± 2.98   \\ \cline{2-8} 
 			& Soft Tissue & 10.28 ± 1.68     & 2.61 ± 1.95      & 17.00 ± 1.91  & 5.75 ± 3.45   & 15.34 ± 1.69  & 2.51 ± 0.99   \\ \cline{2-8} 
 			& Lung        & 6.68 ± 1.58      & 0.73 ± 0.59      & 9.04 ± 1.58   & 0.94 ± 0.81   & 9.08 ± 1.81   & 1.28 ± 0.81   \\ \hline
 		\end{tabular}
 	\end{table}%
 
 	The SPR maps generated using physics-based stoichiometric method from original DECT and sDECT image datasets are shown in Figure 8 as an exemplary case. The SPR of sDECT is close to that of DECT overall. As indicated by arrows, severe noise and artifacts can be seen on SPR of DECT around where noise and artifacts are present in original DECT images, usually around bones and along the longest pathway of x-ray beams, while they are highly mitigated in the SPR by sDECT. The statistical analysis of SPR quality among all patients is summarized in Table 3. The SPR by sDECT predicted by our method is accurate within patient body with an average NMSE of 1.13±0.26\%, an average NCC of 0.96±0.14 and an average PSNR of 40.53±1.39 dB. Among all VOIs except lung, our results have mean error less than 3\%, 17\%-40\% lower noise levels and 15\%-44\% voxels of artifacts than those generated by DECT with statistical significance. Noise and artifacts in lung do not show significant difference between our SPR maps from sDECT and SPR maps from DECT.
 	
 	\begin{figure}
 		\begin{center}
 			\noindent \includegraphics*[width=6.50in, height=4.20in, keepaspectratio=true]{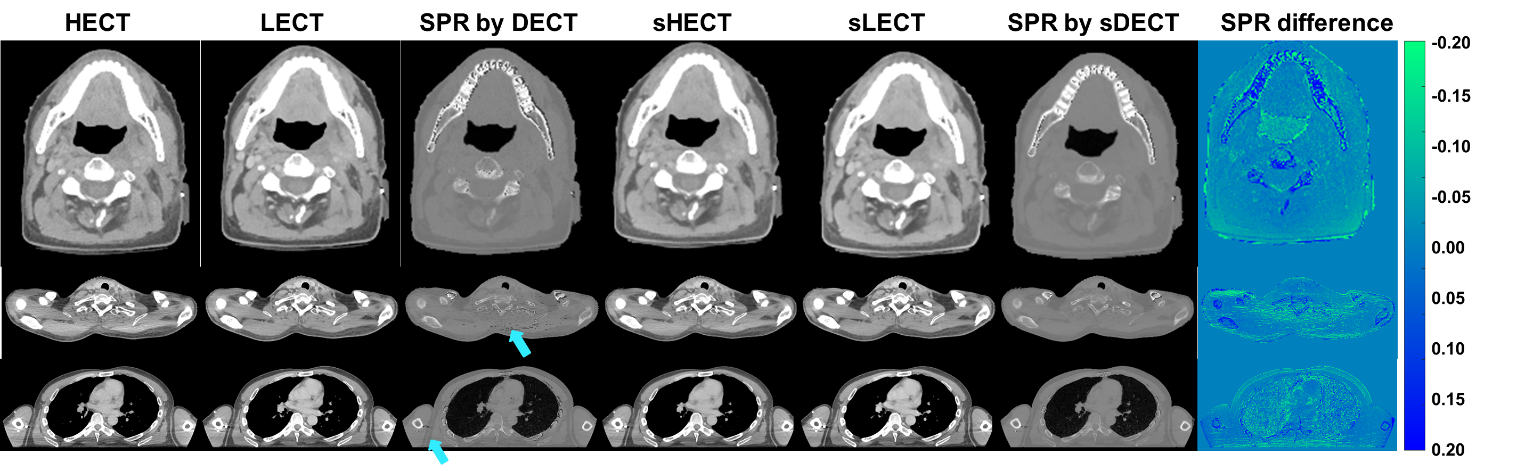}
 		\end{center}
 		
 		\noindent Fig.8 SPR maps generated from DECT and sDECT, and their difference maps (SPR by sDECT - SPR by DECT). Window levels: CT [-200 200], SPR [0 2]. 
 	\end{figure}
 
 	\begin{table}[htbp]
 		\centering
 		\caption{Comparison of proposed method to method A and method B with MAE and ME for four different regions of interest: bone, fat, soft tissue and lung.}
 		\hspace*{-1cm}
 		\begin{tabular}{cccccccc}
 			\hline
 			& NMSE (\%)  & NCC         & PSNR (dB)   &                  &             &            &                   \\ \hline
 			Overall     & 1.13±0.26  & 0.96±0.14   & 40.53±1.39  &                  &             &            &                   \\ \hline
 			&            & \multicolumn{3}{c}{Noise}                    & \multicolumn{3}{c}{\% of voxel in artifacts} \\ \hline
 			VOI         & ME (\%)    & DECT        & sDECT       & P-value          & DECT        & sDECT      & P-value           \\ \hline
 			Lung        & 11.01±4.78 & 0.071±0.015 & 0.073±0.012 & 0.137            & 1.11±0.44   & 1.13±0.38  & 0.883             \\ \hline
 			Adipose     & 2.33±1.50  & 0.041±0.024 & 0.026±0.014 & \textless{}0.001 & 1.30±0.60   & 0.88±0.50  & \textless{}0.001  \\ \hline
 			Soft tissue & 0.55±0.27  & 0.040±0.008 & 0.024±0.004 & \textless{}0.001 & 1.17±0.33   & 1.00±0.29  & 0.041             \\ \hline
 			Bone        & -1.73±2.30 & 0.291±0.107 & 0.240±0.073 & \textless{}0.001 & 1.23±0.90   & 0.68±0.85  & 0.003             \\ \hline
 		\end{tabular}
 	\end{table}%

	\bigbreak
	
	\noindent 
	\section{Discussion}
	
	In this study, we proposed a novel machine-learning based method to synthesize DECT from SECT with a possible application in proton radiation therapy treatment planning. We evaluated the accuracy of synthesized DECT maps using our method in the context of head-and-neck cancer patients. The proposed method successfully generated sLECT and sHECT images from SECT image with an average MAE around 30 HU. The corresponding SPR maps generated from sDECT are quantitatively close to those from original DECT over all, with reduced noise level and artifacts.
	
	Our study is one of the earlier attempts to generate DECT images from SECT images concurrent with methods proposed by Zhao et al. (Method A) \cite{RN44} and Li et al. (Method B) \cite{RN45}. Our study is different from these existing studies in several ways. First, CNN-based networks are used in both of the current studies. Compared with their methods, we proposed a GAN-based network, which uses additional discriminators to enhance the reality of sDECT images, and we added gradient difference error in loss function to increase the similarity of the local tissue structure between sDECT and its training target. Secondly, both of the existing studies implemented their methods in synthesizing HECT from LECT, while our method synthesized both LECT and HECT from SECT. Considering that the current single energy scan in clinic usually has an X-ray energy spectrum falling in between those of LECT and HECT, using SECT to generate the two DECT images, as we did in this study, is more readily to implement in current clinical workflow. Moreover, these studies focus mainly on the quality of sDECT and its diagnostic application such as virtual non-contrast CT and material decomposition images. Our study investigated the feasibility of sDECT in generating SPR for proton therapy treatment planning. Our study also moved a step further in quantifying the noise and artifact performance in the derived SPR maps and demonstrated the superiority of sDECT over DECT. 
	
	In this study, we trained and evaluated our machine learning model on DECT images acquired from TwinBeam. TwinBeam CT has inferior energy separation than other DECT modalities, which leads to a higher sensitivity of SPR maps to artifacts and noise on DECT image datasets \cite{RN1597}. It can be a potential reason that the SPR results generated from DECT images by the physics-based method demonstrated larger error and higher noise level than previously reported \cite{RN35}, where DECT images were acquired by two sequential scans at two different energy levels, which have a larger separation between the two energy spectra than the TwinBeam scan scheme used in this study. However, note that the proposed method does not specify the scan scheme, thus it is applicable to other DECT modalities which have better energy separation for training datasets. 
	
	The SPR maps derived from sDECT show lower noise and artifacts than those derived from original DECT images. This noise robustness feature in SPR maps using sDECT is consistent with other studies \cite{RN45, RN44} where sDECT-derived material decomposition images or virtual non-contrast images are found outperforming those derived from DECT in noise and artifacts. We believe it can be potentially attributed to the strong correlation between the noise in the two sDECT image datasets, and give a brief explanation here. The noise on the derived images is a weighted summation of the noise on HECT and LECT minus a noise correlation term \cite{RN39}. The noise on HECT and LECT is not correlated, thus the noise correlation term is zero. However, sHECT and sLECT is correlated since they are from the same SECT, thus the noise correlation term is no longer zero, which results in a smaller noise on the derived SPR maps. In future study, we will have a detailed examination into the correlation between the noise on sHECT and sLECT generated from machine learning-based method, and a rigorous mathematical proof about its noise propagation to the derived images.
	
	It should be noted that DECT-based SPR maps, although are regarded as clinical gold standard, also contain noise and errors from imaging and modeling uncertainty \cite{RN58}, thus they do not present real physical SPR values. The “error” of SPR presented in this study should be understood as “difference” from DECT-based SPR.
	
	The SPR maps derived from sDECT showed an overestimated bias of 11\% on average over those by acquired DECT in lung region. However, such large discrepancy does not necessarily indicate a large error of SPR by sDECT in lung, because studies reported that the proton SPR of low density materials such as lung would be incorrectly estimated when the physics-based stoichiometric method is used \cite{RN50} on the DECT images. Su et al. reported \cite{RN48} that the SPR estimated using physics-based stoichiometric method in lung was underestimated by around 8\% when comparing with calculation based on chemical compositions. It seems that the error of our SPR results in lung may reduce to about 3\% overestimation over physical reality, while it still needs to be confirmed with potential reasons to be figured out in future study.
	
	In this study, the neural network has shown success in adding new physical information to SECT. Note that such physical information added by neural network did not emerge out of the void but actually from the huge amount of training datasets behind the neural network. In our case, the input SECT does not provide information of energy dependence of attenuation on each pixel as DECT, but the neutral network, after learning the attenuation-energy dependence by seeing enough representative SECT and DECT pairs, can predict this information for a new input SECT. Though the usage of neural networks to generate synthetic DECT images is a tempting opportunity, there are cases, i.e. ability to distinguish between contrastversus dense tumor, where they can fail. However, with further assessments it may be valuable tool for SPR calculation rather than diagnosis. Compared with diagnosis that is focused on the subtle CT appearance, the calculation of SPR and the following proton dose calculation would be more forgiving to local and small errors since they involve the whole imaging area of patients.
	
	Computational cost for training a model is a challenge for deep learning-based methods. We implemented the proposed algorithm with Python 3.7 and TensorFlow as in-house software on an NVIDIA Tesla V100 GPU with 32GB of memory. Adam gradient optimizer with learning rate of 2e-4 was used for optimization. In the present study, the training stage requires ~30 GB and ~19.1 hours for the training datasets of 19 patients, and ~2 minutes for each patient in testing stage.
	
	In this study, we tested our method in the head-and-neck region. Head-and-neck patients feature high anatomical complexity and variability between patients. The tumor shape, size, and location can vary greatly for different patients, and it is common to see the tumor changing the exterior body shape, which is challenging for learning-based method. Future studies should involve a comprehensive evaluation with a larger population of patients with diverse anatomical abnormalities to further reduce bias during the model training. Different testing and training datasets from different institutes would also be valuable to evaluate the clinical utility of our method. Moreover, the proposed method can be applied to other treatment sites of clinical importance for proton therapy, which would be of great interest for expanding this work to the clinic. However, additional sites like abdomen and pelvic region have their inherent challenges associated with DECT image acquisition like organ motions in abdomen or artifacts in overweight patients or patients with hip replacement. In addition to training more models, dosimetric studies need to be carried out to validate the SPR maps generated from sDECT by the proposed method with clinical proton plans.

	\bigbreak
	
	\noindent 
	\section{Conclusions}
	
	We applied a novel deep learning-based approach, namely residual attention GAN, to synthesize sLECT and sHECT images from SECT images for potential applications in the clinic where DECT scanner is not available. The proposed method demonstrated a comparable level of precision in reliably generating synthetic images when compared to ground truth, and noise robustness in derived SPR maps.

	\noindent 
	\bigbreak
	{\bf ACKNOWLEDGEMENT}
	
	This research is supported in part by the National Cancer Institute of the National Institutes of Health under Award Number R01CA215718, and Emory Winship Pilot Grant.

	\noindent 
	\bigbreak
	{\bf Disclosures}
	
	The authors declare no conflicts of interest.

	\noindent 
	
	\bibliographystyle{plainnat}  
	\bibliography{arxiv}      

\begin{thebibliography}{53}
\providecommand{\natexlab}[1]{#1}
\providecommand{\url}[1]{\texttt{#1}}
\expandafter\ifx\csname urlstyle\endcsname\relax
  \providecommand{\doi}[1]{doi: #1}\else
  \providecommand{\doi}{doi: \begingroup \urlstyle{rm}\Url}\fi

\bibitem[RN2()]{RN29}
URL
  \url{https://www.cmsimaging.com/assets/img/brands/hitachi/CT/scenaria128/dualEnergy.pdf}.

\bibitem[RN4(2017)]{RN45}
November 01, 2017 2017.
\newblock URL \url{https://ui.adsabs.harvard.edu/abs/2017arXiv171107118L}.

\bibitem[RN4(2019)]{RN44}
June 01, 2019 2019.
\newblock URL \url{https://ui.adsabs.harvard.edu/abs/2019arXiv190604874Z}.

\bibitem[Almeida et~al.(2017)Almeida, Schyns, Ollers, van Elmpt, Parodi,
  Landry, and Verhaegen]{RN1343}
I.~P. Almeida, L.~E. Schyns, M.~C. Ollers, W.~van Elmpt, K.~Parodi, G.~Landry,
  and F.~Verhaegen.
\newblock Dual-energy ct quantitative imaging: a comparison study between
  twin-beam and dual-source ct scanners.
\newblock \emph{Med Phys}, 44\penalty0 (1):\penalty0 171--179, 2017.
\newblock ISSN 0094-2405.
\newblock \doi{10.1002/mp.12000}.

\bibitem[Alvarez and Macovski(1976)]{RN13}
R.~E. Alvarez and A.~Macovski.
\newblock Energy-selective reconstructions in x-ray computerized tomography.
\newblock \emph{Phys Med Biol}, 21\penalty0 (5):\penalty0 733--44, 1976.
\newblock ISSN 0031-9155 (Print) 0031-9155.
\newblock \doi{10.1088/0031-9155/21/5/002}.

\bibitem[Bar et~al.(2018)Bar, Lalonde, Zhang, Jee, Yang, Sharp, Liu, Royle,
  Bouchard, and Lu]{RN27}
E.~Bar, A.~Lalonde, R.~Zhang, K.~W. Jee, K.~Yang, G.~Sharp, B.~Liu, G.~Royle,
  H.~Bouchard, and H.~M. Lu.
\newblock Experimental validation of two dual-energy ct methods for proton
  therapy using heterogeneous tissue samples.
\newblock \emph{Med Phys}, 45\penalty0 (1):\penalty0 48--59, 2018.
\newblock ISSN 0094-2405.
\newblock \doi{10.1002/mp.12666}.

\bibitem[Bourque et~al.(2014)Bourque, Carrier, and Bouchard]{RN34}
A.~E. Bourque, J.~F. Carrier, and H.~Bouchard.
\newblock A stoichiometric calibration method for dual energy computed
  tomography.
\newblock \emph{Phys Med Biol}, 59\penalty0 (8):\penalty0 2059--88, 2014.
\newblock ISSN 0031-9155.
\newblock \doi{10.1088/0031-9155/59/8/2059}.

\bibitem[Briechle and Hanebeck(2001)]{RN33}
Kai Briechle and Uwe~D. Hanebeck.
\newblock \emph{Template matching using fast normalized cross correlation},
  volume 4387 of \emph{Aerospace/Defense Sensing, Simulation, and Controls}.
\newblock SPIE, 2001.
\newblock URL \url{https://doi.org/10.1117/12.421129}.

\bibitem[Dong et~al.(2019)Dong, Lei, Tian, Wang, Patel, Curran, Jani, Liu, and
  Yang]{RN31}
X.~Dong, Y.~Lei, S.~Tian, T.~Wang, P.~Patel, W.~J. Curran, A.~B. Jani, T.~Liu,
  and X.~Yang.
\newblock Synthetic mri-aided multi-organ segmentation on male pelvic ct using
  cycle consistent deep attention network.
\newblock \emph{Radiother Oncol}, 2019.
\newblock ISSN 0167-8140.
\newblock \doi{10.1016/j.radonc.2019.09.028}.

\bibitem[Engel et~al.(2008)Engel, Herrmann, and Zeitler]{RN53}
Klaus~J. Engel, Christoph Herrmann, and Günter Zeitler.
\newblock X-ray scattering in single- and dual-source ct.
\newblock 35\penalty0 (1):\penalty0 318--332, 2008.
\newblock ISSN 0094-2405.
\newblock \doi{10.1118/1.2820901}.
\newblock URL
  \url{https://aapm.onlinelibrary.wiley.com/doi/abs/10.1118/1.2820901}.

\bibitem[Forghani et~al.(2017)Forghani, De~Man, and Gupta]{RN43}
R.~Forghani, B.~De~Man, and R.~Gupta.
\newblock Dual-energy computed tomography: Physical principles, approaches to
  scanning, usage, and implementation: Part 1.
\newblock \emph{Neuroimaging Clin N Am}, 27\penalty0 (3):\penalty0 371--384,
  2017.
\newblock ISSN 1052-5149.
\newblock \doi{10.1016/j.nic.2017.03.002}.

\bibitem[Glazebrook et~al.(2011)Glazebrook, Guimaraes, Murthy, Black, Bongartz,
  Manek, Leng, Fletcher, and McCollough]{RN11}
K.~N. Glazebrook, L.~S. Guimaraes, N.~S. Murthy, D.~F. Black, T.~Bongartz,
  N.~J. Manek, S.~Leng, J.~G. Fletcher, and C.~H. McCollough.
\newblock Identification of intraarticular and periarticular uric acid crystals
  with dual-energy ct: initial evaluation.
\newblock \emph{Radiology}, 261\penalty0 (2):\penalty0 516--24, 2011.
\newblock ISSN 0033-8419.
\newblock \doi{10.1148/radiol.11102485}.

\bibitem[Goodsitt et~al.(2011)Goodsitt, Christodoulou, and Larson]{RN659}
M.~M. Goodsitt, E.~G. Christodoulou, and S.~C. Larson.
\newblock Accuracies of the synthesized monochromatic ct numbers and effective
  atomic numbers obtained with a rapid kvp switching dual energy ct scanner.
\newblock \emph{Med Phys}, 38\penalty0 (4):\penalty0 2222--32, 2011.
\newblock ISSN 0094-2405 (Print) 0094-2405.

\bibitem[Graser et~al.(2009)Graser, Johnson, Chandarana, and Macari]{RN9}
A.~Graser, T.~R. Johnson, H.~Chandarana, and M.~Macari.
\newblock Dual energy ct: preliminary observations and potential clinical
  applications in the abdomen.
\newblock \emph{Eur Radiol}, 19\penalty0 (1):\penalty0 13--23, 2009.
\newblock ISSN 0938-7994.
\newblock \doi{10.1007/s00330-008-1122-7}.

\bibitem[Hansen et~al.(2015)Hansen, Seco, Sorensen, Petersen, Wildberger,
  Verhaegen, and Landry]{RN25}
D.~C. Hansen, J.~Seco, T.~S. Sorensen, J.~B. Petersen, J.~E. Wildberger,
  F.~Verhaegen, and G.~Landry.
\newblock A simulation study on proton computed tomography (ct) stopping power
  accuracy using dual energy ct scans as benchmark.
\newblock \emph{Acta Oncol}, 54\penalty0 (9):\penalty0 1638--42, 2015.
\newblock ISSN 0284-186x.
\newblock \doi{10.3109/0284186x.2015.1061212}.

\bibitem[Harms et~al.(2019)Harms, Lei, Wang, Zhang, Zhou, Tang, Curran, Liu,
  and Yang]{RN30}
Joseph Harms, Yang Lei, Tonghe Wang, Rongxiao Zhang, Jun Zhou, Xiangyang Tang,
  Walter~J. Curran, Tian Liu, and Xiaofeng Yang.
\newblock Paired cycle-gan-based image correction for quantitative cone-beam
  computed tomography.
\newblock \emph{Med Phys}, 46\penalty0 (9):\penalty0 3998--4009, 2019.
\newblock ISSN 0094-2405.
\newblock \doi{10.1002/mp.13656}.
\newblock URL
  \url{https://aapm.onlinelibrary.wiley.com/doi/abs/10.1002/mp.13656}.

\bibitem[Hunemohr et~al.(2014)Hunemohr, Krauss, Tremmel, Ackermann, Jakel, and
  Greilich]{RN20}
N.~Hunemohr, B.~Krauss, C.~Tremmel, B.~Ackermann, O.~Jakel, and S.~Greilich.
\newblock Experimental verification of ion stopping power prediction from dual
  energy ct data in tissue surrogates.
\newblock \emph{Phys Med Biol}, 59\penalty0 (1):\penalty0 83--96, 2014.
\newblock ISSN 0031-9155.
\newblock \doi{10.1088/0031-9155/59/1/83}.

\bibitem[Jakel et~al.(2001)Jakel, Jacob, Schardt, Karger, and Hartmann]{RN22}
O.~Jakel, C.~Jacob, D.~Schardt, C.~P. Karger, and G.~H. Hartmann.
\newblock Relation between carbon ion ranges and x-ray ct numbers.
\newblock \emph{Med Phys}, 28\penalty0 (4):\penalty0 701--3, 2001.
\newblock ISSN 0094-2405 (Print) 0094-2405.
\newblock \doi{10.1118/1.1357455}.

\bibitem[Johnson(2012)]{RN38}
T.~R. Johnson.
\newblock Dual-energy ct: general principles.
\newblock \emph{AJR Am J Roentgenol}, 199\penalty0 (5 Suppl):\penalty0 S3--8,
  2012.
\newblock ISSN 0361-803x.
\newblock \doi{10.2214/ajr.12.9116}.

\bibitem[Kau et~al.(2011)Kau, Eicher, Reiterer, Niedermayer, Rabitsch, Senft,
  and Hausegger]{RN4}
T.~Kau, W.~Eicher, C.~Reiterer, M.~Niedermayer, E.~Rabitsch, B.~Senft, and
  K.~A. Hausegger.
\newblock Dual-energy ct angiography in peripheral arterial occlusive
  disease-accuracy of maximum intensity projections in clinical routine and
  subgroup analysis.
\newblock \emph{Eur Radiol}, 21\penalty0 (8):\penalty0 1677--86, 2011.
\newblock ISSN 0938-7994.
\newblock \doi{10.1007/s00330-011-2099-1}.

\bibitem[Kyriakou and Kalender(2007)]{RN56}
Y.~Kyriakou and W.~A. Kalender.
\newblock Intensity distribution and impact of scatter for dual-source ct.
\newblock \emph{Phys Med Biol}, 52\penalty0 (23):\penalty0 6969--89, 2007.
\newblock ISSN 0031-9155 (Print) 0031-9155.
\newblock \doi{10.1088/0031-9155/52/23/014}.

\bibitem[Lei et~al.(2019)Lei, Harms, Wang, Liu, Shu, Jani, Curran, Mao, Liu,
  and Yang]{RN32}
Y.~Lei, J.~Harms, T.~Wang, Y.~Liu, H.~K. Shu, A.~B. Jani, W.~J. Curran, H.~Mao,
  T.~Liu, and X.~Yang.
\newblock Mri-only based synthetic ct generation using dense cycle consistent
  generative adversarial networks.
\newblock \emph{Med Phys}, 46\penalty0 (8):\penalty0 3565--3581, 2019.
\newblock ISSN 0094-2405.
\newblock \doi{10.1002/mp.13617}.

\bibitem[Li et~al.(2017)Li, Lee, Duan, Shen, Zhou, Jia, and Yang]{RN58}
B.~Li, H.~C. Lee, X.~Duan, C.~Shen, L.~Zhou, X.~Jia, and M.~Yang.
\newblock Comprehensive analysis of proton range uncertainties related to
  stopping-power-ratio estimation using dual-energy ct imaging.
\newblock \emph{Phys Med Biol}, 62\penalty0 (17):\penalty0 7056--7074, 2017.
\newblock ISSN 0031-9155.
\newblock \doi{10.1088/1361-6560/aa7dc9}.

\bibitem[Marin et~al.(2014)Marin, Boll, Mileto, and Nelson]{RN54}
Daniele Marin, Daniel~T. Boll, Achille Mileto, and Rendon~C. Nelson.
\newblock State of the art: Dual-energy ct of the abdomen.
\newblock 271\penalty0 (2):\penalty0 327--342, 2014.
\newblock \doi{10.1148/radiol.14131480}.
\newblock URL \url{https://pubs.rsna.org/doi/abs/10.1148/radiol.14131480}.

\bibitem[Matsumoto et~al.(2011)Matsumoto, Jinzaki, Tanami, Ueno, Yamada, and
  Kuribayashi]{RN15}
K.~Matsumoto, M.~Jinzaki, Y.~Tanami, A.~Ueno, M.~Yamada, and S.~Kuribayashi.
\newblock Virtual monochromatic spectral imaging with fast kilovoltage
  switching: improved image quality as compared with that obtained with
  conventional 120-kvp ct.
\newblock \emph{Radiology}, 259\penalty0 (1):\penalty0 257--62, 2011.
\newblock ISSN 0033-8419.
\newblock \doi{10.1148/radiol.11100978}.

\bibitem[McCollough et~al.(2015)McCollough, Leng, Yu, and Fletcher]{RN37}
C.~H. McCollough, S.~Leng, L.~Yu, and J.~G. Fletcher.
\newblock Dual- and multi-energy ct: Principles, technical approaches, and
  clinical applications.
\newblock \emph{Radiology}, 276\penalty0 (3):\penalty0 637--53, 2015.
\newblock ISSN 0033-8419.
\newblock \doi{10.1148/radiol.2015142631}.

\bibitem[Mohler et~al.(2018)Mohler, Russ, Wohlfahrt, Elter, Runz, Richter, and
  Greilich]{RN28}
C.~Mohler, T.~Russ, P.~Wohlfahrt, A.~Elter, A.~Runz, C.~Richter, and
  S.~Greilich.
\newblock Experimental verification of stopping-power prediction from single-
  and dual-energy computed tomography in biological tissues.
\newblock \emph{Phys Med Biol}, 63\penalty0 (2):\penalty0 025001, 2018.
\newblock ISSN 0031-9155.
\newblock \doi{10.1088/1361-6560/aaa1c9}.

\bibitem[Nicolaou et~al.(2010)Nicolaou, Yong-Hing, Galea-Soler, Hou, Louis, and
  Munk]{RN10}
S.~Nicolaou, C.~J. Yong-Hing, S.~Galea-Soler, D.~J. Hou, L.~Louis, and P.~Munk.
\newblock Dual-energy ct as a potential new diagnostic tool in the management
  of gout in the acute setting.
\newblock \emph{AJR Am J Roentgenol}, 194\penalty0 (4):\penalty0 1072--8, 2010.
\newblock ISSN 0361-803x.
\newblock \doi{10.2214/ajr.09.2428}.

\bibitem[Niu et~al.(2014)Niu, Dong, Petrongolo, and Zhu]{RN39}
T.~Niu, X.~Dong, M.~Petrongolo, and L.~Zhu.
\newblock Iterative image-domain decomposition for dual-energy ct.
\newblock \emph{Med Phys}, 41\penalty0 (4):\penalty0 041901, 2014.
\newblock ISSN 0094-2405.
\newblock \doi{10.1118/1.4866386}.

\bibitem[Petersilka et~al.(2010)Petersilka, Stierstorfer, Bruder, and
  Flohr]{RN1593}
M.~Petersilka, K.~Stierstorfer, H.~Bruder, and T.~Flohr.
\newblock Strategies for scatter correction in dual source ct.
\newblock \emph{Med Phys}, 37\penalty0 (11):\penalty0 5971--92, 2010.
\newblock ISSN 0094-2405 (Print) 0094-2405.
\newblock \doi{10.1118/1.3504606}.

\bibitem[Petrongolo and Zhu(2015)]{RN1685}
M.~Petrongolo and L.~Zhu.
\newblock Noise suppression for dual-energy ct through entropy minimization.
\newblock \emph{IEEE Trans Med Imaging}, 34\penalty0 (11):\penalty0 2286--97,
  2015.
\newblock ISSN 0278-0062.
\newblock \doi{10.1109/tmi.2015.2429000}.

\bibitem[Primak et~al.(2007)Primak, Fletcher, Vrtiska, Dzyubak, Lieske,
  Jackson, Williams, and McCollough]{RN208}
A.~N. Primak, J.~G. Fletcher, T.~J. Vrtiska, O.~P. Dzyubak, J.~C. Lieske, M.~E.
  Jackson, Jr. Williams, J.~C., and C.~H. McCollough.
\newblock Noninvasive differentiation of uric acid versus non-uric acid kidney
  stones using dual-energy ct.
\newblock \emph{Acad Radiol}, 14\penalty0 (12):\penalty0 1441--7, 2007.
\newblock ISSN 1076-6332 (Print) 1076-6332.
\newblock \doi{10.1016/j.acra.2007.09.016}.

\bibitem[Roele et~al.(2017)Roele, Timmer, Vaassen, van Kroonenburgh, and
  Postma]{RN51}
Elise~D. Roele, Veronique C. M.~L. Timmer, Lauretta A.~A. Vaassen, Anna M.
  J.~L. van Kroonenburgh, and A.~A. Postma.
\newblock Dual-energy ct in head and neck imaging.
\newblock \emph{Current radiology reports}, 5\penalty0 (5):\penalty0 19--19,
  2017.
\newblock ISSN 2167-4825.
\newblock \doi{10.1007/s40134-017-0213-0}.
\newblock URL \url{https://www.ncbi.nlm.nih.gov/pubmed/28435761
  https://www.ncbi.nlm.nih.gov/pmc/articles/PMC5371622/}.

\bibitem[Ruzsics et~al.(2008)Ruzsics, Lee, Zwerner, Gebregziabher, Costello,
  and Schoepf]{RN1}
B.~Ruzsics, H.~Lee, P.~L. Zwerner, M.~Gebregziabher, P.~Costello, and U.~J.
  Schoepf.
\newblock Dual-energy ct of the heart for diagnosing coronary artery stenosis
  and myocardial ischemia-initial experience.
\newblock \emph{Eur Radiol}, 18\penalty0 (11):\penalty0 2414--24, 2008.
\newblock ISSN 0938-7994.
\newblock \doi{10.1007/s00330-008-1022-x}.

\bibitem[Ruzsics et~al.(2009)Ruzsics, Schwarz, Schoepf, Lee, Bastarrika,
  Chiaramida, Costello, and Zwerner]{RN1185}
B.~Ruzsics, F.~Schwarz, U.~J. Schoepf, Y.~S. Lee, G.~Bastarrika, S.~A.
  Chiaramida, P.~Costello, and P.~L. Zwerner.
\newblock Comparison of dual-energy computed tomography of the heart with
  single photon emission computed tomography for assessment of coronary artery
  stenosis and of the myocardial blood supply.
\newblock \emph{Am J Cardiol}, 104\penalty0 (3):\penalty0 318--26, 2009.
\newblock ISSN 0002-9149.
\newblock \doi{10.1016/j.amjcard.2009.03.051}.

\bibitem[Schneider et~al.(1996)Schneider, Pedroni, and Lomax]{RN47}
U.~Schneider, E.~Pedroni, and A.~Lomax.
\newblock The calibration of ct hounsfield units for radiotherapy treatment
  planning.
\newblock \emph{Phys Med Biol}, 41\penalty0 (1):\penalty0 111--24, 1996.
\newblock ISSN 0031-9155 (Print) 0031-9155.
\newblock \doi{10.1088/0031-9155/41/1/009}.

\bibitem[Su et~al.(2018)Su, Kuo, Jordan, Van~Hedent, Klahr, Wei, Al~Helo,
  Liang, Qian, Pereira, Rassouli, Gilkeson, Traughber, Cheng, and Muzic]{RN48}
K.~H. Su, J.~W. Kuo, D.~W. Jordan, S.~Van~Hedent, P.~Klahr, Z.~Wei, R.~Al~Helo,
  F.~Liang, P.~Qian, G.~C. Pereira, N.~Rassouli, R.~C. Gilkeson, B.~J.
  Traughber, C.~W. Cheng, and R.~F. Muzic.
\newblock Machine learning-based dual-energy ct parametric mapping.
\newblock \emph{Phys Med Biol}, 63\penalty0 (12):\penalty0 125001, 2018.
\newblock ISSN 0031-9155.
\newblock \doi{10.1088/1361-6560/aac711}.

\bibitem[Taasti et~al.(2017)Taasti, Michalak, Hansen, Deisher, Kruse, Krauss,
  Muren, Petersen, and McCollough]{RN3010}
V.~T. Taasti, G.~J. Michalak, D.~C. Hansen, A.~J. Deisher, J.~J. Kruse,
  B.~Krauss, L.~P. Muren, J.~B.~B. Petersen, and C.~H. McCollough.
\newblock Validation of proton stopping power ratio estimation based on dual
  energy ct using fresh tissue samples.
\newblock \emph{Phys Med Biol}, 63\penalty0 (1):\penalty0 015012, 2017.
\newblock ISSN 0031-9155.
\newblock \doi{10.1088/1361-6560/aa952f}.

\bibitem[Taasti et~al.(2018)Taasti, Muren, Jensen, Petersen, Thygesen, Tietze,
  Grau, and Hansen]{RN19}
Vicki~Trier Taasti, Ludvig~Paul Muren, Kenneth Jensen, Jørgen Breede~Baltzer
  Petersen, Jesper Thygesen, Anna Tietze, Cai Grau, and David~Christoffer
  Hansen.
\newblock Comparison of single and dual energy ct for stopping power
  determination in proton therapy of head and neck cancer.
\newblock \emph{Physics and Imaging in Radiation Oncology}, 6:\penalty0 14--19,
  2018.
\newblock ISSN 2405-6316.
\newblock \doi{https://doi.org/10.1016/j.phro.2018.04.002}.
\newblock URL
  \url{http://www.sciencedirect.com/science/article/pii/S2405631617300842}.

\bibitem[Thieme et~al.(2012)Thieme, Graute, Nikolaou, Maxien, Reiser, Hacker,
  and Johnson]{RN5}
S.~F. Thieme, V.~Graute, K.~Nikolaou, D.~Maxien, M.~F. Reiser, M.~Hacker, and
  T.~R. Johnson.
\newblock Dual energy ct lung perfusion imaging--correlation with spect/ct.
\newblock \emph{Eur J Radiol}, 81\penalty0 (2):\penalty0 360--5, 2012.
\newblock ISSN 0720-048x.
\newblock \doi{10.1016/j.ejrad.2010.11.037}.

\bibitem[Tran et~al.(2009)Tran, Straka, Roos, Napel, and Fleischmann]{RN202}
D.~N. Tran, M.~Straka, J.~E. Roos, S.~Napel, and D.~Fleischmann.
\newblock Dual-energy ct discrimination of iodine and calcium: experimental
  results and implications for lower extremity ct angiography.
\newblock \emph{Acad Radiol}, 16\penalty0 (2):\penalty0 160--71, 2009.
\newblock ISSN 1076-6332.
\newblock \doi{10.1016/j.acra.2008.09.004}.

\bibitem[van Elmpt et~al.(2016)van Elmpt, Landry, Das, and Verhaegen]{RN17}
W.~van Elmpt, G.~Landry, M.~Das, and F.~Verhaegen.
\newblock Dual energy ct in radiotherapy: Current applications and future
  outlook.
\newblock \emph{Radiother Oncol}, 119\penalty0 (1):\penalty0 137--44, 2016.
\newblock ISSN 0167-8140.
\newblock \doi{10.1016/j.radonc.2016.02.026}.

\bibitem[Vogl et~al.(2012)Vogl, Schulz, Bauer, Stöver, Sader, and
  Tawfik]{RN52}
Thomas~J. Vogl, Boris Schulz, Ralf~W. Bauer, Timo Stöver, Robert Sader, and
  Ahmed~M. Tawfik.
\newblock Dual-energy ct applications in head and neck imaging.
\newblock \emph{American Journal of Roentgenology}, 199\penalty0
  (5supplement):\penalty0 S34--S39, 2012.
\newblock ISSN 0361-803X.
\newblock \doi{10.2214/AJR.12.9113}.
\newblock URL \url{https://doi.org/10.2214/AJR.12.9113}.

\bibitem[Wang and Zhu(2016)]{RN41}
T.~Wang and L.~Zhu.
\newblock Dual energy ct with one full scan and a second sparse-view scan using
  structure preserving iterative reconstruction (spir).
\newblock \emph{Phys Med Biol}, 61\penalty0 (18):\penalty0 6684--6706, 2016.
\newblock ISSN 0031-9155.
\newblock \doi{10.1088/0031-9155/61/18/6684}.

\bibitem[Wang et~al.(2019)Wang, Ghavidel, Beitler, Tang, Lei, Curran, Liu, and
  Yang]{RN1597}
T.~Wang, B.~B. Ghavidel, J.~J. Beitler, X.~Tang, Y.~Lei, W.~J. Curran, T.~Liu,
  and X.~Yang.
\newblock Optimal virtual monoenergetic image in "twinbeam" dual-energy ct for
  organs-at-risk delineation based on contrast-noise-ratio in head-and-neck
  radiotherapy.
\newblock \emph{J Appl Clin Med Phys}, 20\penalty0 (2):\penalty0 121--128,
  2019.
\newblock ISSN 1526-9914.
\newblock \doi{10.1002/acm2.12539}.

\bibitem[Wang et~al.(2018)Wang, Yu, Wang, Zu, Lalush, Lin, Wu, Zhou, Shen, and
  Zhou]{RN1821}
Y.~Wang, B.~Yu, L.~Wang, C.~Zu, D.~S. Lalush, W.~Lin, X.~Wu, J.~Zhou, D.~Shen,
  and L.~Zhou.
\newblock 3d conditional generative adversarial networks for high-quality pet
  image estimation at low dose.
\newblock \emph{Neuroimage}, 174:\penalty0 550--562, 2018.
\newblock ISSN 1053-8119.
\newblock \doi{10.1016/j.neuroimage.2018.03.045}.

\bibitem[Warren et~al.(2015)Warren, Partridge, Hill, and Peach]{RN50}
D.~R. Warren, M.~Partridge, M.~A. Hill, and K.~Peach.
\newblock Improved calibration of mass stopping power in low density tissue for
  a proton pencil beam algorithm.
\newblock \emph{Phys Med Biol}, 60\penalty0 (11):\penalty0 4243--61, 2015.
\newblock ISSN 0031-9155.
\newblock \doi{10.1088/0031-9155/60/11/4243}.

\bibitem[Watanabe et~al.(2009)Watanabe, Uotani, Nakazawa, Higashi, Yamada,
  Hori, Kanzaki, Fukuda, Itoh, and Naito]{RN3}
Y.~Watanabe, K.~Uotani, T.~Nakazawa, M.~Higashi, N.~Yamada, Y.~Hori,
  S.~Kanzaki, T.~Fukuda, T.~Itoh, and H.~Naito.
\newblock Dual-energy direct bone removal ct angiography for evaluation of
  intracranial aneurysm or stenosis: comparison with conventional digital
  subtraction angiography.
\newblock \emph{Eur Radiol}, 19\penalty0 (4):\penalty0 1019--24, 2009.
\newblock ISSN 0938-7994.
\newblock \doi{10.1007/s00330-008-1213-5}.

\bibitem[Yang et~al.(2010)Yang, Virshup, Clayton, Zhu, Mohan, and Dong]{RN1592}
M.~Yang, G.~Virshup, J.~Clayton, X.~R. Zhu, R.~Mohan, and L.~Dong.
\newblock Theoretical variance analysis of single- and dual-energy computed
  tomography methods for calculating proton stopping power ratios of biological
  tissues.
\newblock \emph{Phys Med Biol}, 55\penalty0 (5):\penalty0 1343--62, 2010.
\newblock ISSN 0031-9155.
\newblock \doi{10.1088/0031-9155/55/5/006}.

\bibitem[Yu et~al.(2011)Yu, Christner, Leng, Wang, Fletcher, and
  McCollough]{RN660}
L.~Yu, J.~A. Christner, S.~Leng, J.~Wang, J.~G. Fletcher, and C.~H. McCollough.
\newblock Virtual monochromatic imaging in dual-source dual-energy ct:
  radiation dose and image quality.
\newblock \emph{Med Phys}, 38\penalty0 (12):\penalty0 6371--9, 2011.
\newblock ISSN 0094-2405 (Print) 0094-2405.
\newblock \doi{10.1118/1.3658568}.

\bibitem[Yu et~al.(2012)Yu, Leng, and McCollough]{RN12}
L.~Yu, S.~Leng, and C.~H. McCollough.
\newblock Dual-energy ct-based monochromatic imaging.
\newblock \emph{AJR Am J Roentgenol}, 199\penalty0 (5 Suppl):\penalty0 S9--s15,
  2012.
\newblock ISSN 0361-803x.
\newblock \doi{10.2214/ajr.12.9121}.

\bibitem[Zhang et~al.(2013)Zhang, Zhou, Schoepf, Sheng, Wu, Krazinski,
  Silverman, Meinel, Zhao, Zhang, and Lu]{RN6}
Long~Jiang Zhang, Chang~Sheng Zhou, U.~Joseph Schoepf, Hui~Xue Sheng,
  Sheng~Yong Wu, Aleksander~W. Krazinski, Justin~R. Silverman, Felix~G. Meinel,
  Yan~E. Zhao, Zong~Jun Zhang, and Guang Ming 
\newblock Dual-energy ct lung ventilation/perfusion imaging for diagnosing
  pulmonary embolism.
\newblock \emph{Eur Radiol}, 23\penalty0 (10):\penalty0 2666--2675, 2013.
\newblock ISSN 1432-1084.
\newblock \doi{10.1007/s00330-013-2907-x}.
\newblock URL \url{https://doi.org/10.1007/s00330-013-2907-x}.

\bibitem[Zhu and Penfold(2016)]{RN35}
J.~Zhu and S.~N. Penfold.
\newblock Dosimetric comparison of stopping power calibration with dual-energy
  ct and single-energy ct in proton therapy treatment planning.
\newblock \emph{Med Phys}, 43\penalty0 (6):\penalty0 2845--2854, 2016.
\newblock ISSN 0094-2405.
\newblock \doi{10.1118/1.4948683}.

\end{thebibliography}
	
\end{document}